\documentclass[10pt, conference, compsocconf]{IEEEtran}

\usepackage{pgfplots}
\usepackage{algpseudocode}
\usepackage{algorithm}
\usepackage{enumerate}
\usepackage{cite}
\usepackage{amsfonts}
\usepackage[stable]{footmisc}

\begin{document}

\title{A Brief History of Web Crawlers}

\author{\IEEEauthorblockN{Seyed M. Mirtaheri, Mustafa Emre Din\c{c}t\"{u}rk, Salman Hooshmand, Gregor V. Bochmann, Guy-Vincent Jourdan}
\IEEEauthorblockA{School of Electrical Engineering and Computer Science\\
University of Ottawa\\
Ottawa, Ontario, Canada\\
\{smirt016,mdinc075, shooshmand\}@uottawa.ca, \{bochmann,gvj\}@eecs.uottawa.ca}\\
\IEEEauthorblockN{Iosif Viorel Onut}
\IEEEauthorblockA{Security AppScan\textsuperscript{\textregistered} Enterprise, IBM\\
770 Palladium Dr\\
Ottawa, Ontario, Canada\\
vioonut@ca.ibm.com}
}

\maketitle

\begin{abstract}

Web crawlers have a long and interesting history. 
Early web crawlers collected statistics about the web. In addition to collecting statistics about the web and indexing the applications for search engines, 
modern crawlers can be used to perform accessibility and vulnerability checks on the application. 

Quick expansion of the web, and the complexity added to web applications have made the process of crawling a very challenging one. 
Throughout the history of web crawling many researchers and industrial groups 
addressed different issues and challenges that web crawlers face. Different solutions have been proposed to reduce the time and cost of crawling. 
Performing an exhaustive crawl is a challenging question. 
Additionally, capturing the model of a modern web application and extracting data from it automatically is another open question. 

What follows is a brief history of different technique and algorithms used from the early days of crawling up to 
the recent days. We introduce criteria to evaluate the relative performance of web crawlers. 
Based on these criteria we plot the evolution of web crawlers and compare their performance.

\end{abstract}

\section{Introduction}
Crawling is the process of exploring web applications automatically. 
The web crawler aims at discovering the web pages of a web application by navigating through the application. 
This is usually done by simulating the possible user interactions considering just the client-side of the application. 

As the amount of information on the web has been increasing drastically, web users increasingly rely on search engines to find desired data. 
In order for search engines to learn about the new data as it becomes available on the web, the web crawler has to constantly crawl and update the search 
engine database.
 
We start by introducing motivations of crawling, defining the problem of crawling formally, quick overview of history crawlers and requirements of a good crawler first.

\subsection{Motivations for Crawling}
There are several important motivations for crawling. 
The main three motivations are: 

\begin{itemize} 
	\item Content indexing for search engines. Every search engine requires a web crawler to fetch the data from the web. 
	\item Automated testing and model checking of the web application
	\item Automated security testing and vulnerability assessment. 
		Many web applications use sensitive data and provide critical services.
		To address the security concerns for web applications, many commercial and open-source automated web application security scanners
		have been developed. These tools aim at detecting possible issues, 
		such as security vulnerabilities and usability issues, in an automated and efficient manner\cite{bau2010state, doupe2010johnny}.
		They require a web crawler to discover the states of the application scanned. 
\end{itemize}	

\subsection{The Evolution of Web Crawlers}

In the literature on web-crawling, a web crawler is basically a software that starts from a set of seed URLs and downloads all the web pages associated with these URLs. After fetching a web page associated with a URL, the URL is removed from the working queue. The web crawler then parses the downloaded page, extracts the linked URLs form it, and adds new URLs to the list of seed URLs. This process continues iteratively until all of the contents reachable from seed URLs are reached. 

The traditional definition of a web crawler assumes that all the content of a web application is reachable through URLs. Soon in the history of web crawling it became clear that such web crawlers can not deal with the complexities added by interactive web applications that rely on the user input to generate web pages. This scenario often arises when the web application is an interface to a database and it relies on user input to retrieve contents from the database. The new field of \textit{Deep Web-Crawling} was born to address this issue. 

Availability of powerful client-side web-browsers, as well as the wide adaptation to technologies such as HTML5 and AJAX, gave birth to a new pattern in designing web applications called \textit{Rich Internet Application} (RIA). RIAs move part of the computation from the server to the client. This new pattern of designing web applications led to complex client side applications that increased the speed and interactivity of the web application, while it reduced the network traffic per request. 

Despite many added values, RIAs introduced some unique challenges to web crawlers. In a RIA, user interaction often results in execution of client side \textit{events}. Execution of an event in a RIA often changes the state of the web application on the client side, which is represented in the form of a \textit{Document Object Model} (DOM)\cite{domBook}. This change in the state of DOM does not necessarily means changing the URL. Traditional web crawlers rely heavily on the URL and changes to the DOM that do not alter the URL are invisible to them. Although deep web crawling increased the ability of the web crawlers to retrieve data from web applications, it fails to address changes to DOM that do not affect the URL. The new and recent field of \textit{RIA web-crawling} attempts to address the problem of RIA crawling.

\subsection{Problem Definition}
\label{probdef}

A web application can be modeled as a directed graph, and the \textit{World Wide Web} can be modeled as a forest of such graphs. The problem of Web crawling is the problem of discovering all the nodes in this forest. In the application graph, each node represents a state of the application and each edge a transition from one state to another. 

As web applications evolved, the definition of the state of the application evolved as well. In the context of traditional web applications, states in the application graph are pages with distinct URLs and edges are hyperlinks between pages i.e. there exist an edge between two nodes in the graph if there exist a link between the two pages. In the context of deep web crawling, transitions are constructed based on users input. This is in contrast with hyperlink transitions which always redirect the application to the same target page. In a deep web application, any action that causes submission of a form is a possible edge in the graph. 
 
In RIAs, the assumption that pages are nodes in the graph is not valid, since the client side code can change the application state without changing the page URL. 
Therefore nodes here are application states denoted by their DOM, and edges are not restricted to forms that submit elements, since each element can communicate with the server and partially update the current state. Edges, in this context, are client side actions (e.g. in JavaScript) assigned to DOM elements and can be detected by web crawler. Unlike the traditional web applications where jumps to arbitrary states are possible, in a RIA, the execution of sequence of events from the current state or from a seed URL is required to reach a particular state. 

\begin{table*}
	\centering
	\begin{tabular}{|p{1.5cm}|p{4cm}|p{11cm}|}
		\hline
			\parbox{1.5cm}{\vspace{1 mm}\centering \textbf{Web crawler type} \vspace{1 mm}} & 
			\parbox{4cm}{\vspace{1 mm}\centering \textbf{Input} \vspace{1 mm}} & 
			\parbox{11cm}{\vspace{1 mm}\centering \textbf{Application graph components} \vspace{1 mm}}\\
		\hline
		\hline
			\parbox{1.5cm}{\vspace{1 mm} Traditional \vspace{1 mm} } &
			\parbox{4cm}{\vspace{1 mm} Set of seed URLs \vspace{1 mm} } &
			\parbox{11cm}{\vspace{1 mm} Nodes are pages with distinct URL and a directed edge exist from page $p_1$ to page $p_2$ if there is a hyperlink in page $p_1$ that points to page $p_2$ \vspace{1 mm} }\\
		\hline
			\parbox{1.5cm}{\vspace{1 mm} Deep \vspace{1 mm} } &
			\parbox{4cm}{\vspace{1 mm} Set of Seed URLs, user context specific data, domain taxonomy \vspace{1 mm} } &
			\parbox{11cm}{\vspace{1 mm} Nodes are pages and a directed edge exists between page $p_1$ to page $p_2$ if submitting a form in page $p_1$ gets the user to page $p_2$. \vspace{1 mm} }\\
		\hline
			\parbox{1.5cm}{\vspace{1 mm} RIA \vspace{1 mm} } &
			\parbox{4cm}{\vspace{1 mm} A starting page \vspace{1 mm} } &
			\parbox{11cm}{\vspace{1 mm} Nodes are DOM states of the application and a directed edge exist from DOM $d_1$ to DOM $d_2$ if there is a client-side JavaScript event, detectable by the web crawler, that if triggered on $d_1$ changes the DOM state to $d_2$ \vspace{1 mm} }\\
		\hline
		\hline
			\parbox{1.5cm}{\vspace{1 mm} Unified Model \vspace{1 mm} } &
			\parbox{4cm}{\vspace{1 mm} A seed URL \vspace{1 mm} } &
			\parbox{11cm}{\vspace{1 mm} Nodes are calculated based on DOM and the URL. An edge is a transmission between two states triggered through client side events. Redirecting the browser is a special client side event. \vspace{1 mm} }\\
		\hline

	\end{tabular} 
	\caption{Different categories of web crawlers}
	\label{tableStages}
\end{table*}

The three models can be unified by defining the state of the application based on the state of the DOM as well as other parameters such as the page URL, rather than the URL or the DOM alone. A hyperlink in a traditional web application does not only change the page URL, but it also changes the state of the DOM. In this model changing the page URL can be viewed as a special client side event that updates the entire DOM. Similarly, submission of a HTML form in a deep web application leads to a particular state of DOM once the response comes back from the server. In both cases the final DOM states can be used to enumerate the states of the application. Table \ref{tableStages} summarizes different categories of web crawlers. 

\subsection{Requirements}
\label{secRequirements}

Several design goals have been considered for web crawlers. \textit{Coverage} and \textit{freshness} are among the first \cite{OlstonN10}. Coverage measures the relative number of pages discovered by the web crawler. Ideally given enough time the web crawler has to find all pages and build the complete model of the application. This property is referred to as \textit{Completeness}. Coverage captures the static behaviour of traditional web applications well. It may fail, however, to capture the performance of the web crawler in crawling dynamically created web pages. The search engine index has to be updated constantly to reflect changes in web pages created dynamically. The ability of the web crawler to retrieve latest updates is measured through \textit{freshness}. 

An important and old issue in designing web crawlers is called \textit{politeness}\cite{Heydon99mercator:a}. Early web crawlers had no mechanism to stop them from bombing a server with many requests. As the result while crawling a website they could have lunched an inadvertent \textit{Denial of Service}(DoS) attack and exhaust the target server resources to the point that it would interrupt normal operation of the server. Politeness was the concept introduced to put a cap on the number of requests sent to a web-server per unit of time. A polite web crawler avoids launching an inadvertent DoS attack on the target server. Another old problem that web crawlers faced are \textit{traps}. Traps are seemingly large set of websites with arbitrary data that are meant to waste the web crawler resources. Integration of \textit{black-lists} allowed web crawlers to avoid traps. Among the challenges web crawlers faced in the mid 90s was \textit{scalability}\cite{burner1997crawling}. Throughout the history of web-crawling, the exponential growth of the web and its constantly evolving nature has been hard to match by web crawlers. In addition to these requirements, the web crawler's model of application should be \textit{correct} and reflect true content and structure of the application. 

In the context of deep-web crawling Raghavan and Garcia-Molina\cite{RaghavanCrawlingTheHiddenWeb} suggest two more requirements. In this context, \textit{Submission efficiency} is defined as the ratio of submitted forms leading to result pages with new data; and \textit{Lenient submission efficiency} measures if a form submission is semantically correct (e.g., submitting a company name as input to a form element that was intended to be an author name)

In the context of RIA crawling a non-functional requirement considered by Kamara et al.\cite{ICWE2011} called \textit{efficiency}. Efficiency means discovering valuable information as soon as possible. For example states are more important than transitions and should be found first instead of finding transitions leading to already known states. This is particularly important if the web crawler will perform a partial crawl rather than a full crawl. 

This paper defines web crawling and its requirements, and based on the defined model classifies web crawlers. 

A brief history of traditional web crawlers\footnote{See Olston and Najork\cite{OlstonN10} for a survey of traditional web crawlers.}, deep web crawlers\footnote{See He et al.\cite{AccessingTheDeepWebASurvey} for a survey of deep web crawlers.}, and RIA crawlers\footnote{See Choudhary et al.\cite{stateOfArt} for a survey of RIA crawlers.} is presented in sections II-IV. Based on this brief history and the model defined, taxonomy of web crawling is then presented in section V. Section VI concludes the paper with some open questions and future works in web crawling. 

\section{Crawling Traditional Web Applications}
\label{tiaCrawling}

Web crawlers were written as early as 1993. This year gave birth to four web crawlers: \textit{World Wide Web Wanderer}, \textit{Jump Station}, \textit{World Wide Web Worm}\cite{McBryan94}, and \textit{RBSE spider}. These four spiders mainly collected information and statistic about the web using a set of seed URLs. Early web crawlers iteratively downloaded URLs and updated their repository of URLs through the downloaded web pages. 

The next year, 1994, two new web crawlers appeared: \textit{WebCrawler} and \textit{MOMspider}. In addition to collecting stats and data about the state of the web, these two web crawlers introduced concepts of \textit{politeness} and \textit{black-lists} to traditional web crawlers. \textit{WebCrawler} is considered to be the first parallel web crawler by downloading 15 links simultaneously. From \textit{World Wide Web Worm} to \textit{WebCrawler}, the number of indexed pages increased from 110,000 to 2 million. Shortly after, in the coming years a few commercial web crawlers became available: \textit{Lycos}, \textit{Infoseek}, \textit{Excite}, \textit{AltaVista} and \textit{HotBot}.

In 1998, Brin and Page\cite{Brin:1998} tried to address the issue of scalability by introducing a large scale web crawler called \textit{Google}. Google addressed the problem of scalability in several ways: Firstly it leveraged many low level optimizations to reduce disk access time through techniques such as compression and indexing. Secondly, and on a higher level, Google calculated the probability of a user visiting a page through an algorithm called \textit{PageRank}. PageRange calculates the probability of a user visiting a page by taking into account the number of links that point to the page as well as the style of those links. Having this probability, Google simulated an arbitrary user and visited a page as often as the user did. Such approach optimizes the resources available to the web crawler by reducing the rate at which the web crawler visits unattractive pages. Through this technique, Google achieved high \textit{freshness}. Architecturally, Google used a master-slave architecture with a master server (called \textit{URLServer}) dispatching URLs to a set of slave nodes. The slave nodes retrieve the assigned pages by downloading them from the web. At its peak, the first implementation of Google reached 100 page downloads per second. 

The issue of scalability was further addressed by Allan Heydon and Marc Najork in a tool called \textit{Mercator}\cite{Heydon99mercator:a} in 1999. Additionally Mercator attempted to address the problem of extendability of web crawlers. To address extensibility it took advantage of a modular Java-based framework. This architecture allowed third-party components to be integrated into Mercator. To address the problem of scalability, Mercator tried to solve the problem of \textit{URL-Seen}. The URL-Seen problem answers the question of whether or not a URL was seen before. This seemingly trivial problem gets very time-consuming as the size of the URL list grows. Mercator increased the scalability of URL-Seen by batch disk checks. In this mode hashes of discovered URLs got stored in RAM. When the size of these hashes grows beyond a certain limit, the list was compared against the URLs stored on the disk, and the list itself on the disk was updated. Using this technique, the second version of Mercator crawled 891 million pages. Mercator got integrated into \textit{AltaVista} in 2001. 

IBM introduced \textit{WebFountain}\cite{Edwards01anadaptive} in 2001. WebFountain was a fully distributed web crawler and its objective was not only to index the web, but also to create a local copy of it. This local copy was \textit{incremental} meaning that a copy of the page was kept indefinitely on the local space, and this copy got updated as often as WebFountain visited the page. In WebFountain, major components such as the scheduler were distributed and the crawling was an ongoing process where the local copy of the web only grew. These features, as well as deployment of efficient technologies such as the \textit{Message Passing Interface} (MPI), made WebFountain a scalable web crawler with high freshness rate. In a simulation, WebFountain managed to scale with a growing web. This simulated web originally had 500 million pages and it grew to twice its size every 400 days. 

In 2002, \textit{Polybot}\cite{Shkapenyuk02designand} addressed the problem of URL-Seen scalability by enhancing the batch disk check technique. Polybot used Red-Black tree to keep the URLs and and when the tree grows beyond a certain limit, it was merged with a sorted list in main memory. Using this data structure to handle URL-Seen test, Polybot managed to scan 120 million pages. In the same year, \textit{UbiCrawler}\cite{ubicrawl} dealt with the problem of URL-Seen with a different, more peer-to-peer (P2P), approach. UbiCrawler used consistent hashing to distribute URLs among web crawler nodes. In this model no centralized unit calculates whether or not a URL was seen before, but when a URL is discovered it is passed to the node responsible to answer the test. The node responsible to do this calculation is detected by taking the hash of the URL and map it to the list of nodes. With five 1GHz PCs and fifty threads, UbiCrawler reached a download rate of 10 million pages per day. 

In addition to Polybot and UbiCrawler, in 2002 Tang et al. introduced \textit{pSearch}\cite{Tang02psearch}. pSearch uses two algorithms called \textit{P2P Vector Space Model (pVSM)} and \textit{P2P Latent Semantic Indexing (pLSI)} to crawl the web on a P2P network. VSM and LSI in turn use vector representation to calculate the relation between queries and the documents. Additionally pSearch took advantage of \textit{Distributed Hash Tables} (DHT) routing algorithms to address scalability.

Two other studies used DHTs over P2P networks. In 2003, Li et al\cite{noTheFeasibilityofP2P} used this technique to scale up certain tasks such as clustering of contents and bloom filters. In 2004, Loo et al~\cite{distributedWebCrawlingDHTs} addressed the question of scalability of web crawlers and used the technique to partition URLs among the crawlers. One of the underlying assumption in this work is the availability of high speed communication medium. The implemented prototype requested 800,000 pages from more than 70,000 web crawlers in 15 minutes. 

In 2005, Exposto et al.\cite{Exposto:2005} augmented partitioning of URLs among a set of crawling nodes in a P2P architecture by taking into account servers geographical information. Such an augmentation reduced the overall time of the crawl by allocating target servers to a node physically closest to them. 

In 2008, an extremely scalable web crawler called \textit{IRLbot} ran for 41.27 days on a quad-CPU AMD Opteron 2.6 GHz server and it crawled over 6.38 billion web pages\cite{Lee_irlbot:scaling}. IRLbot primarily addressed the \textit{URL-Seen} problem by breaking it down into three sub-problems: \textsc{check}, \textsc{update} and \textsc{check+update}. To address these sub-problems, IRLbot introduced a framework called \textit{Disk Repository with Update Management} (DRUM). DRUM optimizes disk access by segmenting the disk into several \textit{disk buckets}. For each disk bucket, DRUM also allocates a corresponding bucket on the RAM. Each URL is mapped to a bucket. At first a URL was stored in its RAM bucket. Once a bucket on the RAM is fulled, the corresponding disk bucket is accessed in batch mode. This batch mode access, as well as the two-stage bucketing system used, allowed DRUM to store large number of URLs on the disk such that its performance would not degrade as the number of URLs increases.

\section{Crawling Deep Web}
\label{deepWebCrawling}

As server-side programming and scripting languages, such as PHP and ASP, got momentum, more and more databases became accessible online through interacting with a web application. The applications often delegated creation and generation of contents to the executable files using \textit{Common Gateway Interface} (CGI). In this model, programmers often hosted their data on databases and used HTML forms to query them. Thus a web crawler can not access all of the contents of a web application merely by following hyperlinks and downloading their corresponding web page. These contents are \textit{hidden} from the web crawler point of view and thus are referred to as \textit{deep web}\cite{AccessingTheDeepWebASurvey}. 

In 1998, Lawrence and Giles\cite{SearchingTheWorldWideWeb} estimated that 80 percent of web contents were hidden in 1998. Later in 2000, BrightPlanet suggested that the deep web contents is 500 times larger than what surfaces through following hyperlinks (referred to as \textit{shallow web})\cite{BrightPlanet}. The size of the deep web is rapidly growing as more companies are moving their data to databases and set up interfaces for the users to access them\cite{BrightPlanet}. 

Only a small fraction of the deep web is indexed by search engines. In 2007, He et al\cite{AccessingTheDeepWebASurvey} randomly sampled one million IPs and crawled these IPs looking for deep webs through HTML form elements. The study also defined a depth factor from the original seed IP address and constrained itself to depth of three. Among the sampled IPs, 126 deep web sites were found. These deep websites had 406 query gateways to 190 databases. Based on these results with 99 percent confidence interval, the study estimates that at the time of that writing, there existed $1,097,000$ to $1,419,000$ database query gateways on the web. The study further estimated that Google and Yahoo search engines each has visited only 32 percent of the deep web. To make the matters worst the study also estimated that 84 percent of the covered objects overlap between the two search engines, so combining the discovered objects by the two search engines does not increase the percentage of the visited deep web by much. 

The second generation of web crawlers took the deep web into account. Information retrieval from the deep web meant interacting with HTML forms. To retrieve information hidden in the deep web, the web crawler would submit the HTML form many times, each time filled with a different dataset. Thus the problem of crawling the deep web got reduced to the problem of assigning proper values to the HTML form fields. 

The open and difficult question to answer in designing a deep web crawler is how to meaningfully assign values to the fields in a query form\cite{SiphoningHiddenWebData}. As Barbosa and Freire\cite{SiphoningHiddenWebData} explain, it is easy to assign values to fields of certain types such as radio buttons. The difficult field to deal with, however, is text box inputs. Many different proposals tried to answer this question: 

\begin{itemize} 
	\item In 2001, Raghavan and Garcia-Molina\cite{RaghavanCrawlingTheHiddenWeb} proposed a method to fill up text box inputs that mostly depend on human output. 
	\item In 2002, Liddle et al.\cite{ExtractingDataBehindWebForms} described a method to detect form elements and fabricate a HTTP GET and POST request using default values specified for each field. The proposed algorithm is not fully automated and asks for user input when required. 
	\item In 2004, Barbosa and Freire\cite{SiphoningHiddenWebData} proposed a two phase algorithm to generate textual queries. The first stage collected a set of data from the website and used that to associate weights to keywords. The second phase used a greedy algorithm to retrieve as much contents as possible with minimum number of queries. 
	\item In 2005, Ntoulas et al.\cite{downloadingTextualHiddenWeb} further advanced the process by defining three policies for sending queries to the interface: a random policy, a policy based on the frequency of keywords in a reference document, and an adaptive policy that learns from the downloaded pages. Given four entry points, this study retrieved 90 percent of the deep web with only 100 requests. 
	\item In 2008, Lu et al.\cite{WebIntelligenceIntelligentAgentTechnology} map the problem of maximizing the coverage per number of requests to the problem of \textit{set-covering}\cite{clrs} and uses a classical approach to solve this problem. 
\end{itemize}

\section{Crawling Rich Internet Applications}
\label{riaCrawling}

Powerful client side browsers and availability of client-side technologies lead to a shift in computation from server-side to the client-side. This shift of computation, also creates contents that are often hidden from traditional web-crawlers and are referred to as "Client-side hidden-web''\cite{MesbahHiddenWeb2013}. In 2013, Behfarshad and Mesbah studies 500 web-sites and found that 95 percent of the subject websites contain client-side hidden-web, and among the 95 percent web-sites, 62 percent of the application states are considered client-side hidden-web. Extrapolating these numbers puts almost 59 percent of the web contents at the time of this writing as client-side hidden-web.

RIA crawling differs from traditional web application crawling in several frontiers. Although limited, there has been some research focusing on crawling of RIAs. One of the earliest attempts to crawl RIAs is by Duda et al in 2007\cite{ETHajaxcrawl,ETHcrawlmaster, ETHindexmaster}. This work presents a working prototype of a RIA crawler that indexed RIAs using a Breath-First-Search algorithm. In 2008, Mesbah et al. introduced \textit{Crawljax}\cite{crawljaxconf, crawljaxjournal} a RIA crawler that took the user-interface into account and used the changes made to the user interface to direct the crawling strategy. Crawljax aimed at crawling and taking a static snapshot of each AJAX state for indexing and testing. In the same year, Amalfitano et al.\cite{amalreverse, amalriatest, amalexperimenting, amaltechniques} addressed automatic testing of RIAs using execution traces obtained from AJAX applications. 

This section surveys different aspects of RIA crawling. Different strategies can be used to choose an unexecuted events to execute. Different strategies effect how early the web crawler finds new states and the overall time of crawling. Section \ref{crawlStrategies} surveys some of the strategies studied in recent years. Section \ref{domEquivalency} explains different approaches to determine if two DOMs are equivalent. Section \ref{parallelCrawling} surveys parallelism and concurrency for RIA crawling. Automated testing and ranking algorithms are explored in Sections \ref{automatedTesting} and \ref{pageRank}, respectively. 

\subsection{Crawling Strategy}
\label{crawlStrategies}

Until recent years, there has not been much attention on the efficiency requirement, and existing approaches often use either Breadth-First or a Depth-First crawling strategy. 
Duda et al.\cite{ETHajaxcrawl,ETHcrawlmaster, ETHindexmaster} used Breadth-First crawling strategy. As an optimization, the communication cost was reduced by caching the JavaScript function calls (together with actual parameters) that resulted in AJAX requests and the response received from the server. Crawljax \cite{crawljaxconf, crawljaxjournal} extracted a model of the application using a variation of the Depth-First strategy. Its default strategy only explored a subset of the events in each state. This strategy explored an event only from the state where the event was first encountered. The event was not explored on the subsequently discovered states. This default strategy may not find all the states, since executing the same event from different states may lead to different states. However, Crawljax can also be configured to explore all enabled events in each state, in that case its strategy becomes the standard Depth-First crawling strategy.

Amalfitano et al.\cite{amalreverse, amalriatest, amalexperimenting} focused on modelling and testing RIAs using execution traces. The initial work \cite{amalreverse} was based on obtaining execution traces from user-sessions (a manual method). Once the traces are obtained, they are analyzed and an FSM model is formed by grouping together the equivalent user interfaces according to an equivalence relation. In a later paper\cite{amalriatest} CrawlRIA was introduced which automatically generated execution traces using a Depth-First strategy. Starting from the initial state, CrawlRIA executed events in a depth-first manner until a DOM instance that is equivalent to a previously visited DOM instance was reached. Then the sequence of states and events was stored as a trace in a database, and after a reset, crawling continued from the initial state to record another trace. These automatically generated traces were later used to form an FSM model using the same technique that is used in \cite{amalreverse} for user-generated traces.

In 2011, Kamara et al.\cite{ICWE2011, Kamarathesis} present the initial version of the first model-based crawling strategy: the \textit{Hypercube strategy}. The strategy makes predictions by initially assuming the model of the application to be a hypercube structure. The initial implementation had performance drawbacks which prevented the strategy to be practical even when the number of events in the initial state are as few as twenty. These limitation was later removed\cite{ICWE2011}. 

In 2012, Choudhary et al.\cite{Suryathesis} introduced another model-based strategy called the \textit{Menu strategy}. This strategy is optimized for the applications that have the same event always leading to the same state, irrelevant of the source state. Dincruk et al.\cite{emre12n2} introduced a statistical model-based strategy. This strategy uses statistics to determine which events have a high probability to lead to a new stete. 

In the same year, Peng et al.\cite{greedystrategy} suggested to use a \textit{greedy strategy}. In the greedy strategy if there is an un-executed event in the current state (i.e. the state which the web crawler's DOM structure represents) the event is executed. If the current state has no unexplored event, the web crawler transfers to the closest state with an unexecuted event. Two other variants of the greedy strategy are introduced by the authors as well. In these variations, instead of the closest state, the most recently discovered state and the state closest to the initial state are chosen when there is no event to explore in the current state. They experimented with this strategy on simple test applications using different combinations of navigation styles to navigate a sequence of ordered pages. The navigation styles used are previous and next events, events leading to a few of the preceding and succeeding pages from the current page, as well as the events that lead to the first and last page. They concluded that all three variations of the strategy have similar performance in terms of the total number of event executions to finish crawling.

In 2013, Milani Fard and Mesbah\cite{geedyMesbah2013} introduce \textit{FeedEx}: a greedy algorithm to partially crawl a RIAs. 
FeedEx differs from Peng et al.\cite{greedystrategy} in that: Peng et al.\cite{greedystrategy} use a greedy algorithm in finding the closest unexecuted event, whereas, FeedEx defines a matrix to measure the impact of an event and its corresponding state on the crawl. 
The choices are then sorted and the most impactful choice will be executed first. Given enough time, FeedEx will discover entire graph of the application. 

FeedEx defines the impact matrix as a weighted sum of the following four factors:
\begin{itemize}
	\item Code coverage impact: how much of the application code is being executed. 
	\item Navigational diversity: how diversely the crawler explores the application graph.
	\item Page structural diversity: how newly discovered DOMs differ from those already discovered. 
	\item Test model size: the size of the created test model. 
\end{itemize}. 

In the test cases studied, Milani Fard and Mesbah\cite{geedyMesbah2013} show that FeedEx beats three other strategies of Breadth-First search, Depth-First search, and random strategy, in the above-mentioned four factors.

\subsection{DOM Equivalence and Comparison}
\label{domEquivalency}

In the context of traditional web applications it is trivial to determine whether two states are equal: compare their URLs. This problem is not as trivial in the context of RIAs. Different chains of events may lead to the same states with minor differences that do not effect the functionality of the state. Different researchers address this issue differently. Duda et al.\cite{ETHajaxcrawl,ETHcrawlmaster, ETHindexmaster} used equality as the DOM equivalence method. Two states compared based on ``the hash value of the full serialized DOM"\cite{ETHindexmaster}. As admitted in \cite{ETHindexmaster} this equality is too strict and may lead to too many states being produced.

Crawljax \cite{crawljaxconf} used an edit distance (the number of operations that is needed to change one DOM instance to the other, the so-called Levenstein distance) to decide if the current DOM instance corresponds to a different state than the previous one. If the distance is below a certain threshold the current DOM instance is considered equivalent to the previous one. Otherwise, the current DOM instance is hashed and its hash value is compared to the hash values of the already discovered states. Since the notion of distance is not transitive, it is not an equivalence relation in the mathematical sense. For this reason, using a distance has the problem of incorrectly grouping together client-states whose distance is actually above the given threshold. 

In a later paper\cite{crawljaxjournal}, Crawljax improves its DOM equivalency: To decide if a new state is reached, the current DOM instance is compared with all the previously discovered states' DOMs using the mentioned distance heuristic. If the distance of the current DOM instance from each seen DOM instance is above the threshold, then the current DOM is considered as a new state. Although this approach solves the mentioned problem with the previous approach, this method may not be as efficient since it requires to store the DOM-trees and compute the distance of the current DOM to all the discovered DOMs.

Amalfitano et al.\cite{amalexperimenting} proposed DOM equivalence relations based on comparing the set of elements in the two DOM instances. According to this method, two DOM instances are equivalent if both contain the same set of elements. This inclusion is checked based on the indexed paths of the elements, event types and event handlers of the elements. They have also introduced two variations of this relation. In the first variation only visible elements are considered, in the other variation, the index requirement for the paths is removed. 

In 2013, Lo et al.\cite{imagen} in a tool called \textit{Imagen}, consider the problem of transferring a JavaScript session between two clients. 
Imagen improves the definition of client-side state by adding the following items: 
\begin{itemize}
	\item JavaScript functions closure: JavaScript functions can be created dynamically, and their scope is determined at the time of creation. 
	\item JavaScript event listeners: JavaScript allows the programmer to register event-handlers. 
	\item HTML5 elements: Certain elements such as \textit{Opaque Objects} and \textit{Stream Resources}.
\end{itemize}
These items are not ordinarily stored in DOM. Imagen uses code instrumenting and other techniques to add the effect of these features to the state of the application. 

\subsection{Parallel Crawling}
\label{parallelCrawling}

To the best of our knowledge, at the time of this writing only one distributed RIA crawling algorithm exists. Mirtaheri et al.\cite{KitsCrawl} used the JavaScript events to partition the search space and crawl a RIA in parallel. Each web crawler, running on a separate computer, visits all application states, but only executes a subset of the JavaScript events in each state. If execution of an event leads to the discovery of a new state, the information about the new state is propagated to all the web crawlers. Together all the web crawlers cover all JavaScript events in all application states. The proposed algorithm is implemented and evaluated with 15 computers and a satisfactory speedup is demonstrated. Apart from this work, two algorithms are proposed to achieve a degree of concurrency: 

\begin{itemize}

\item In \cite{ETHcrawlmaster}, the authors propose to use multiple web crawlers on RIAs (or on Web crawling) that use hyperlinks together with events for navigation. The suggested method first applies traditional crawling to find the URLs in the application. After traditional crawling terminates, the set of discovered URLs are partitioned and assigned to event-based crawling processes that run independent of each other using their Breadth-First strategy. Since each URL is crawled independently, there is no communication between the web crawlers.

\item Crawljax\cite{crawljaxjournal} uses multiple threads for speeding up event-based crawling of a single URL application. The crawling process starts with a single thread (that uses depth-first strategy). When a thread discovers a state with more than one event, new threads are initiated that will start the exploration from the discovered state and follow one of the unexplored events from there. 

\end{itemize}

\subsection{Automated Testing}
\label{automatedTesting}

Automated testing of RIAs is an important aspect of RIA crawling.  In 2008, Marchetto et al.\cite{statebasedtest} used a state-based testing approach based on a FSM model of the application. The introduced model construction method used static analysis of the JavaScript code and dynamic analysis of user session traces. Abstraction of the DOM states was used rather than the DOM states directly in order to reduce the size of the model. This optimization may require a certain level of manual interaction to ensure correctness of the algorithm. The introduced model produced test sequences that contained \textit{semantically interacting events}\footnote{Two events are semantically interacting if their execution order changes the outcome.}. In 2009, Marchetto and Tonella\cite{searchbasedtest} proposed search-based test sequence generation using hill-climbing rather than exhaustively generating all the sequences up to some maximum length.    

In 2009 and 2010, Crawljax introduced three mechanisms to automate testing of RIAs: Using \textit{invariant-based} testing\cite{crawljaxinvarianttest}, security testing of interactions among web widgets \cite{crawljaxregression}, and regression testing of AJAX applications\cite{crawljaxregression}. 

In 2010, Amalfitono et al.\cite{amalriatest} compared the effectiveness of methods based on execution traces (user generated, web crawler generated and combination of the two) and existing test case reduction techniques based on measures such as state coverage, transition coverage and detecting JavaScript faults. In another study\cite{amaltechniques}, authors used invariant-based testing approach to detect faults visible on the user-interface (invalid HTML, broken links, unsatisfied accessibility requirements) in addition to JavaScript faults (crashes) which may not be visible on the user-interface, but cause faulty behaviour.

\subsection{Ranking (Importance Metric)}
\label{pageRank}

Unlike traditional web application crawling, there has been a limited amount of research in ranking states and pages in the context of RIA crawling. In 2007, Frey\cite{ETHindexmaster} proposed a ranking mechanism for the states in RIAs. The proposed mechanism, called \textit{AjaxRank}, ordered search results by assigning an importance value to states. AjaxRank can be viewed as an adaptation of the PageRank\cite{pagerank}. Similar to PageRank, AjaxRank is connectivity-based but instead of hyperlinks the transitions are considered. In the AjaxRank, the initial state of the URL is given more importance (since it is the only state reachable from anywhere directly), hence the states that are closer to the initial state also get higher ranks.

\section{Taxonomy and Evolution of Web Crawlers}

The wide variety of web crawlers available are designed with different goals in mind. This section classifies and cross-measures the functionalities of different web crawlers based on the design criteria introduced in Section \ref{secRequirements}. It also sketches out a rough architecture of web crawlers as they evolve. Sections \ref{taxoTrad}, \ref{taxoDeep} and \ref{taxoRIA} explain the taxonomy of traditional, deep, and RIA web crawlers, respectively. 

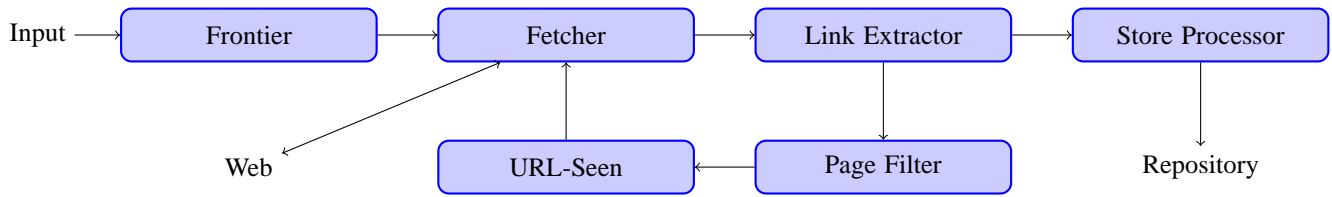
\begin{figure*}
	\begin{center}
		\begin{tikzpicture}
			[auto,block/.style ={rectangle, draw=blue, thick, fill=blue!20, text width=90pt,align=center, rounded corners, minimum height=20pt}]

			\draw (0pt,100pt) node[] (I){Input};	
			\draw (80pt,100pt) node[block] (F){Frontier};
			\draw (80pt,50pt) node[] (W){Web};
			\draw (200pt,100pt) node[block] (E){Fetcher};
			\draw (200pt,50pt) node[block] (U){URL-Seen};
			\draw (320pt,50pt) node[block] (P){Page Filter};
			\draw (320pt,100pt) node[block] (L){Link Extractor};
			\draw (440pt,100pt) node[block] (S){Store Processor};
			\draw (440pt,50pt) node[] (R){Repository};

			\draw[->] (I) -- (F);			
			\draw[->] (F) -- (E);
			\draw[->] (E) -- (L);
			\draw[->] (L) -- (S);

			\draw[<->] (W) -- (E);
			\draw[->] (U) -- (E);
			\draw[->] (L) -- (P);
			\draw[->] (S) -- (R);

			\draw[->] (P) -- (U);
							
		\end{tikzpicture}
	\end{center}

	\caption{Architecture of a Traditional Web Crawler.}
	\label{figArchTrad}

\end{figure*}

\begin{table*}
	\centering
	\begin{tabular}{|p{2.5cm}|p{2.5cm}|p{8cm}|p{3cm}|}
		\hline
			\parbox{2.5cm}{\vspace{1 mm}\centering \textbf{Study} \vspace{1 mm}} & 
			\parbox{2.5cm}{\vspace{1 mm} \textbf{Component} \vspace{1 mm}} & 
			\parbox{8cm}{\vspace{1 mm} \textbf{Method} \vspace{1 mm}} & 
			\parbox{3cm}{\vspace{1 mm} \textbf{Goal}  \vspace{1 mm}  \vspace{1 mm} }\\
		\hline
		\hline
			\parbox{2.5cm}{\vspace{1 mm} \centering WebCrawler \\ MOMspider\cite{OlstonN10} \vspace{1 mm} } &
			\parbox{2.5cm}{\vspace{1 mm} Fetcher \\ Frontier \\ Page filter \vspace{1 mm} } &
			\parbox{8cm}{ \vspace{1 mm} Parallel downloading of 15 links \\ robots.txt \\ Black-list \vspace{1 mm} } &
			\parbox{3cm}{\vspace{1 mm} Scalability \\ Politeness  \vspace{1 mm} }\\
		\hline
			\parbox{2.5cm}{\vspace{1 mm} \centering Google\cite{Brin:1998} \vspace{1 mm}} &
			\parbox{2.5cm}{\vspace{1 mm} Store processor \\ Frontier \vspace{1 mm}} &
			\parbox{8cm}{\vspace{1 mm} Reduce disk access time by compression \\ PageRank \vspace{1 mm}} &
			\parbox{3cm}{\vspace{1 mm} Scalability \\ Coverage \\ Freshness \vspace{1 mm} }\\
		\hline
			\parbox{2.5cm}{\vspace{1 mm} \centering Mercator\cite{Heydon99mercator:a} \vspace{1 mm}} &
			\parbox{2.5cm}{\vspace{1 mm} URL-Seen\vspace{1 mm}} &
			\parbox{8cm}{ \vspace{1 mm}  Batch disk checks and cache \vspace{1 mm}} &
			\parbox{3cm}{\vspace{1 mm}  Scalability \vspace{1 mm} }\\
		\hline
			\parbox{2.5cm}{\vspace{1 mm} \centering WebFountain\cite{Edwards01anadaptive} \vspace{1 mm}} &
			\parbox{2.5cm}{\vspace{1 mm} Storage processor \\ Frontier \\ Fetch \vspace{1 mm}} &
			\parbox{8cm}{ \vspace{1 mm} Local copy of the fetched pages \\ Adaptive download rate \\ Homogenous cluster as hardware \vspace{1 mm}} &
			\parbox{3cm}{\vspace{1 mm} Completeness \\ Freshness \\ Scalability  \vspace{1 mm} }\\
		\hline
			\parbox{2.5cm}{\vspace{1 mm} \centering Polybot\cite{Shkapenyuk02designand} \vspace{1 mm}} &
			\parbox{2.5cm}{\vspace{1 mm} URL-Seen \vspace{1 mm}} &
			\parbox{8cm}{ \vspace{1 mm} Red-Black tree to keep the URLs \vspace{1 mm}} &
			\parbox{3cm}{\vspace{1 mm} Scalability  \vspace{1 mm} }\\
		\hline
			\parbox{2.5cm}{\vspace{1 mm} \centering UbiCrawler\cite{ubicrawl} \vspace{1 mm}} &
			\parbox{2.5cm}{\vspace{1 mm} URL-Seen \vspace{1 mm}} &
			\parbox{8cm}{ \vspace{1 mm} P2P architecture \vspace{1 mm}} &
			\parbox{3cm}{\vspace{1 mm} Scalability  \vspace{1 mm} }\\
		\hline
			\parbox{2.5cm}{\vspace{1 mm} \centering pSearch\cite{Tang02psearch} \vspace{1 mm}} &
			\parbox{2.5cm}{\vspace{1 mm} Store processor \vspace{1 mm}} &
			\parbox{8cm}{ \vspace{1 mm} Distributed Hashing Tables (DHT) \vspace{1 mm}} &
			\parbox{3cm}{\vspace{1 mm} Scalability  \vspace{1 mm} }\\
		\hline
			\parbox{2.5cm}{\vspace{1 mm} \centering Exposto et al.\cite{Exposto:2005} \vspace{1 mm}} &
			\parbox{2.5cm}{\vspace{1 mm} Frontier \vspace{1 mm}} &
			\parbox{8cm}{ \vspace{1 mm} Distributed Hashing Tables (DHT) \vspace{1 mm}} &
			\parbox{3cm}{\vspace{1 mm} Scalability  \vspace{1 mm} }\\
		\hline
			\parbox{2.5cm}{\vspace{1 mm} \centering IRLbotpages\cite{Lee_irlbot:scaling} \vspace{1 mm}} &
			\parbox{2.5cm}{\vspace{1 mm} URL-Seen \vspace{1 mm}} &
			\parbox{8cm}{ \vspace{1 mm} Access time reduction by disk segmentation \vspace{1 mm}} &
			\parbox{3cm}{\vspace{1 mm} Scalability \vspace{1 mm} }\\
		\hline

	\end{tabular} 
	\caption{Taxonomy of Traditional Web Crawlers}
	\label{tabTaxoTrad}
\end{table*}

\subsection{Traditional Web Crawlers}
\label{taxoTrad}

Figure \ref{figArchTrad} shows the architecture of a typical traditional web crawler. In this model \textit{Frontier} gets a set of seed URLs. The seed URLs are passed to a module called \textit{Fetcher} that retrieves the contents of the pages associated with the URLs from the web. These contents are passed to the \textit{Link Extractor}. The latter parses the HTML pages and extract new links from them. Newly discovered links are passed to \textit{Page Filter} and \textit{Store Processor}. Store Processor interacts with the database and stores the discovered links. Page Filter filters URLs that are not interesting to the web crawler. The URLs are then passed to \textit{URL-Seen} module. This module finds the new URLs that are not retrieved yet and passes them to Fetcher for retrieval. This loop continues until all the reachable links are visited. 

Table \ref{tabTaxoTrad} summarizes the design components, design goals and different techniques used by traditional web crawlers. 

\begin{figure*}
	\begin{center}
		\begin{tikzpicture}
			[auto,block/.style ={rectangle, draw=blue, thick, fill=blue!20, text width=90pt,align=center, rounded corners, minimum height=20pt}]

			\draw (0pt,100pt) node[] (I){Input};	
			\draw (80pt,100pt) node[block] (S){Select Fillable};
			\draw (200pt,100pt) node[block] (D){Domain Finder};
			\draw (320pt,100pt) node[block] (U){Submitter};
			\draw (80pt,50pt) node[] (R){Repository};
			\draw (200pt,50pt) node[block] (A){Response Analyzer};
			\draw (320pt,50pt) node[] (W){Web};

			\draw[->] (I) -- (S);
			\draw[->] (S) -- (D);
			\draw[->] (D) -- (U);

			\draw[->] (R) -- (S);
			\draw[->] (R) -- (D);
			\draw[->] (A) -- (R);
			\draw[->] (U) -- (A);
			\draw[<->] (W) -- (U);
				
		\end{tikzpicture}
	\end{center}

	\caption{Architecture of a Deep Web Crawler.}
	\label{figArchDeep}

\end{figure*}

\begin{table*}
	\centering
	\begin{tabular}{|p{2.5cm}|p{2.5cm}|p{8cm}|p{3cm}|}
		\hline
			\parbox{2.5cm}{\vspace{1 mm} \centering \textbf{Study} \vspace{1 mm}} & 
			\parbox{2.5cm}{\vspace{1 mm} \centering \textbf{Component} \vspace{1 mm}} & 
			\parbox{8cm}{\vspace{1 mm} \centering \textbf{Method} \vspace{1 mm}} & 
			\parbox{3cm}{\vspace{1 mm} \centering \textbf{Goal} \vspace{1 mm}}\\
		\hline
		\hline
			\parbox{2.5cm}{\vspace{1 mm} \centering HiWe\cite{RaghavanCrawlingTheHiddenWeb} \vspace{1 mm} } &
			\parbox{2.5cm}{\vspace{1 mm} 
				Select fillable \\ 
				Domain Finder \\ 
				Submitter \\ 
				Response Analyst \vspace{1 mm} } &
			\parbox{8cm}{\vspace{1 mm}  
				Partial page layout and visual adjacency \\ 
				Normalization by stemming etc \\ Approximation matching \\ 
				Manual domain \\ 
				Ignore submitting small or incomplete forms \\ 
				Hash of visually important parts of the page to detect errors\vspace{1 mm} } &
			\parbox{3cm}{\vspace{1 mm} Lenient submission efficiency \\ Submission efficiency  \vspace{1 mm} }\\
		\hline
			\parbox{2.5cm}{\vspace{1 mm} \centering Liddle et al\cite{ExtractingDataBehindWebForms} \vspace{1 mm} } &
			\parbox{2.5cm}{\vspace{1 mm} Select fillable \\ Domain Finder \vspace{1 mm} } &
			\parbox{8cm}{\vspace{1 mm}  
				Fields with finite set of values, ignores automatic filling of text field \\
				Stratified Sampling Method (avoid queries biased toward certain fields) \\
				Detection of new forms inside result page, Removal of repeated form \\
				Concatenation of pages connected through navigational elements  \\
				Stop queries by observing pages with repetitive partial results \\
				Detect record boundaries and computes hash values for each sentence\vspace{1 mm} } &
			\parbox{3cm}{\vspace{1 mm} 
				Lenient submission efficiency \\
				Submission efficiency} \\
		\hline
			\parbox{2.5cm}{\vspace{1 mm} \centering Barbosa and Freire\cite{SiphoningHiddenWebData} \vspace{1 mm} } &
			\parbox{2.5cm}{\vspace{1 mm} Select fillable \\ Domain Finder \\Response Analysis \vspace{1 mm} } &
			\parbox{8cm}{\vspace{1 mm} 
				Single keyword-based queries \\
				Based on collection data associate weights to keywords and uses greedy algorithms to retrieve as much contents with minimum number of queries. \\
				Considers adding stop-words to maximize coverage \\
				Issue queries using dummy words to detect error pages \vspace{1 mm} } &
			\parbox{3cm}{\vspace{1 mm} Lenient submission efficiency \\ Submission efficiency \vspace{1 mm} }\\
		\hline
			\parbox{2.5cm}{\vspace{1 mm} \centering Ntoulas et al\cite{downloadingTextualHiddenWeb} \vspace{1 mm} } &
			\parbox{2.5cm}{\vspace{1 mm} Select fillable \\ Domain Finder \vspace{1 mm} } &
			\parbox{8cm}{\vspace{1 mm} 
				Single-term keyword-based queries \\
				Three policies: random, based on the frequency of keyword in a corpus, and an Adaptive policy that learn from the downloaded pages. 
				maximizing the unique returns of each query\vspace{1 mm} } &
			\parbox{3cm}{\vspace{1 mm} Lenient submission efficiency \\ Submission efficiency \vspace{1 mm} }\\
		\hline
			\parbox{2.5cm}{\vspace{1 mm} \centering Lu et al\cite{WebIntelligenceIntelligentAgentTechnology}\vspace{1 mm} } &
			\parbox{2.5cm}{\vspace{1 mm} Select fillable \\ Domain Finder \vspace{1 mm} } &
			\parbox{8cm}{\vspace{1 mm} 
				querying textual data sources, \\
				Works on sample that represents the original data source. \\
				Maximizing the coverage per number of requests to the problem of set-covering problem\vspace{1 mm} } &
			\parbox{3cm}{\vspace{1 mm} 
				Lenient submission efficiency \\
				Scalability \\
				Submission efficiency \vspace{1 mm} }\\
		\hline

	\end{tabular} 
	\caption{Taxonomy of Deep Web Crawlers}
	\label{tabTaxoDeep}
\end{table*}

\subsection{Deep Web Crawlers}
\label{taxoDeep}

Figure \ref{figArchDeep} shows the architecture of a typical deep web crawler. In this model \textit{Select Fillable} gets as input set of seed URLs, domain data, and user specifics. \textit{Select Fillable} then chooses the HTML elements to interact with. \textit{Domain Finder} uses these data to fill up the HTML forms and passes the results to \textit{Submitter}. Submitter submits the form to the server and retrieves the newly formed page. \textit{Response Analyser} parses the page and, based on the result, updates the repository; and the process continues. 

Table \ref{tabTaxoDeep} summarizes the design components, design goals and different techniques used by deep web crawlers. 

\begin{figure*}
	\begin{center}
		\begin{tikzpicture}
			[auto,block/.style ={rectangle, draw=blue, thick, fill=blue!20, text width=90pt,align=center, rounded corners, minimum height=20pt}]

			\draw (0pt,100pt) node[] (I){Input};	
			\draw (80pt,100pt) node[block] (J){JS-Engine};
			\draw (200pt,100pt) node[block] (D){DOM-Seen};
			\draw (320pt,100pt) node[block] (E){Event Extractor};
			\draw (200pt,50pt) node[block] (S){Strategy};
			\draw (80pt,50pt) node[] (W){Web};
			\draw (320pt,50pt) node[] (M){Model};

			\draw[->] (I) -- (J);
			\draw[->] (J) -- (D);
			\draw[->] (D) -- (E);

			\draw[->] (E) -- (S);
			\draw[->] (S) -- (M);
			\draw[->] (S) -- (J);
			
			\draw[<->] (W) -- (J);

		\end{tikzpicture}
	\end{center}

	\caption{Architecture of a Deep Web Crawler.}
	\label{figArchRIA}

\end{figure*}

\begin{table*}
	\centering
	\begin{tabular}{|p{2.5cm}|p{2.5cm}|p{8cm}|p{3cm}|}
		\hline
			\parbox{2.5cm}{\vspace{1 mm} \centering \textbf{Study} \vspace{1 mm}} & 
			\parbox{2.5cm}{\vspace{1 mm} \centering \textbf{Component} \vspace{1 mm}} & 
			\parbox{8cm}{\vspace{1 mm} \centering \textbf{Method} \vspace{1 mm}} & 
			\parbox{3cm}{\vspace{1 mm} \centering \textbf{Goal}  \vspace{1 mm} }\\
		\hline
		\hline
			\parbox{2.5cm}{\vspace{1 mm} \centering Duda et al\cite{ETHajaxcrawl,ETHcrawlmaster, ETHindexmaster} \vspace{1 mm} } &
			\parbox{2.5cm}{\vspace{1 mm} Strategy \\ JS-Engine \\ DOM-Seen \vspace{1 mm} } &
			\parbox{8cm}{ \vspace{1 mm}
				Breadth-First-Search \\
				Caching the JavaScript function calls and results \\
				Comparing Hash value of the full serialized DOM\vspace{1 mm} \vspace{1 mm}} &
			\parbox{3cm}{\vspace{1 mm} Completeness \\ Efficiency \vspace{1 mm}}\\
		\hline
			\parbox{2.5cm}{\vspace{1 mm} \centering Mesbah et al\cite{crawljaxconf, crawljaxjournal} \vspace{1 mm} } &
			\parbox{2.5cm}{\vspace{1 mm} Strategy \\ DOM-Seen \vspace{1 mm} } &
			\parbox{8cm}{ \vspace{1 mm}
				Depth-First-Search \\
				Explores an event only once \\ 
				New threads are initiated for unexplored events \\
				Comparing Edit distance with all previous states \vspace{1 mm} } &
			\parbox{3cm}{\vspace{1 mm} 
				Completeness \\
				State Coverage Efficiency \\
				Scalability \vspace{1 mm} }\\
		\hline
			\parbox{2.5cm}{\vspace{1 mm} \centering CrawlRIA\cite{amalreverse, amalriatest, amalexperimenting, amaltechniques} \vspace{1 mm} } &
			\parbox{2.5cm}{\vspace{1 mm} Strategy \\ DOM-Seen \vspace{1 mm} } &
			\parbox{8cm}{ \vspace{1 mm}
				Depth-First strategy (Automatically generated using execution traces) \\
				Comparing the set of elements, event types, event handlers in two DOMs\vspace{1 mm} } &
			\parbox{3cm}{\vspace{1 mm} Completeness \vspace{1 mm} }\\
		\hline
			\parbox{2.5cm}{\vspace{1 mm} \centering Kamara et al\cite{Kamarathesis, ICWE2011} \vspace{1 mm} } &
			\parbox{2.5cm}{\vspace{1 mm}  Strategy\vspace{1 mm} } &
			\parbox{8cm}{ \vspace{1 mm}
				Assuming hypercube model for the application.
				Using Minimum Chain Decomposition and Minimum Transition Coverage\vspace{1 mm} } &
			\parbox{3cm}{\vspace{1 mm} State Coverage Efficiency \vspace{1 mm} }\\
		\hline
			\parbox{2.5cm}{\vspace{1 mm} \centering M-Crawler\cite{emre12n1} \vspace{1 mm} } &
			\parbox{2.5cm}{\vspace{1 mm} Strategy \vspace{1 mm} } &
			\parbox{8cm}{ \vspace{1 mm}
				Menu strategy which categorizes events after first two runs \\
				Events which always lead to the same/current state has less priority \\
				Using Rural-Postman solver to explore unexecuted events efficiently\vspace{1 mm} } &
			\parbox{3cm}{\vspace{1 mm} State Coverage Efficiency \\ Completeness \vspace{1 mm} }\\
		\hline
			\parbox{2.5cm}{\vspace{1 mm} \centering Peng et al.\cite{greedystrategy} \vspace{1 mm} } &
			\parbox{2.5cm}{\vspace{1 mm} Strategy \vspace{1 mm} } &
			\parbox{8cm}{ \vspace{1 mm}Choose an event from current state then from the closest state\vspace{1 mm} } &
			\parbox{3cm}{\vspace{1 mm} State Coverage Efficiency \vspace{1 mm} }\\
		\hline
			\parbox{2.5cm}{\vspace{1 mm} \centering AjaxRank\cite{ETHindexmaster} \vspace{1 mm} } &
			\parbox{2.5cm}{\vspace{1 mm} Strategy \\ DOM-Seen \vspace{1 mm} } &
			\parbox{8cm}{ \vspace{1 mm}
				The initial state of the URL is given more importance \\ 
				Similar to PageRank, connectivity-based but instead of hyperlinks the transitions are considered hash value of the content and structure of the DOM\vspace{1 mm} } &
			\parbox{3cm}{\vspace{1 mm} State Coverage Efficiency \vspace{1 mm} }\\
		\hline
			\parbox{2.5cm}{\vspace{1 mm} \centering Dincturk et al.\cite{ICWE2012} \vspace{1 mm} } &
			\parbox{2.5cm}{\vspace{1 mm} Strategy \vspace{1 mm} } &
			\parbox{8cm}{ \vspace{1 mm}
				Considers probability of discovering new 'state' by an event and cost of following the path to event’s state\vspace{1 mm} } &
			\parbox{3cm}{\vspace{1 mm} State Coverage Efficiency \vspace{1 mm} }\\
		\hline
			\parbox{2.5cm}{\vspace{1 mm} \centering Dist-RIA Crawler\cite{KitsCrawl} \vspace{1 mm} } &
			\parbox{2.5cm}{\vspace{1 mm} Strategy \vspace{1 mm} } &
			\parbox{8cm}{ \vspace{1 mm} Uses JavaScript events to partition the search space and run the crawl in parallel on multiple nodes\vspace{1 mm} } &
			\parbox{3cm}{\vspace{1 mm} Scalability  \vspace{1 mm} }\\
		\hline
			\parbox{2.5cm}{\vspace{1 mm} \centering Feedex\cite{geedyMesbah2013} \vspace{1 mm} } &
			\parbox{2.5cm}{\vspace{1 mm} Strategy \vspace{1 mm} } &
			\parbox{8cm}{ \vspace{1 mm} Prioritize events based on their possible impact of the DOM. Considers factors like code coverage, navigational and page structural diversity
\vspace{1 mm} } &
			\parbox{3cm}{\vspace{1 mm} State Coverage Efficiency  \vspace{1 mm} }\\
		\hline	
		
	\end{tabular} 
	\caption{Taxonomy of RIA Web Crawlers}
	\label{tabTaxoRIA}
\end{table*}

\subsection{RIA Web Crawlers}
\label{taxoRIA}

Figure \ref{figArchRIA} shows the architecture of a typical RIA web crawler. \textit{JS-engine} starts a virtual browser and runs a JavaScript engine. It then retrieves the page associated with a seed URL and loads it in the virtual browser. The constructed DOM is passed to the \textit{DOM-Seen} module to determine if this is the first time the DOM is seen. If so, the DOM is passed to \textit{Even Extractor} to extract the JavaScript events form it. The events are then passed to the \textit{Strategy} module. This module decides which event to execute. The chosen event is passed to JS-Engine for further execution. This process continues until all reachable states are seen. 

Table \ref{tabTaxoRIA} summarizes the design components, design goals and different techniques used by RIA web crawlers.

\section{Some Open Questions in Web-Crawling}
\label{conc}

In this paper, we have surveyed the evolution of crawlers, namely traditional, Deep and RIA crawlers. We identified
several design goals and components of each category and developed a taxonomy that classifies different cases of
crawlers accordingly. Traditional web crawling and its scalability has been the topic of extensive research. Similarly, deep-web crawling was addressed in great details. RIA crawling, however, is a new and open area for research. Some of the open questions in the field of RIA crawling are the following: 
\begin{itemize}

\item Model Based Crawling: The problem of designing an efficient strategy for crawling a RIA can be mapped to a graph exploration problem. The objective of the algorithm is to visit every node at least once in an unknown directed graph by minimizing the total sum of the weights of the edges traversed. The offline version of this problem, where the graph is known beforehand, is called the Asymmetric Traveling Salesman Problem (ATSP) which is NP-Hard. Although there are some approximation algorithms for different variations of the unknown graph exploration problem \cite{unknowngraphundirected1,unknowngraphundirected2, unknowngraphdirectedenddpoint, unknowngraphundirectednoendpoint}, not knowing the graph ahead of the time is a major obstacle to deploy these algorithms to crawl RIAs. 

\item Scalability: Problems such as URL-Seen may not exist in the context of RIA crawling. However, a related problem is the \textit{State-Seen} problem: If a DOM state was seen before. 

\item Widget Detection: In order to avoid state explosion, it is crucial to detect independent parts of the interface in a RIA. This can effect ranking of different states, too. 

\end{itemize}
In addition, combining different types of crawlers to build a unified crawler seems another promising research area.

\bibliographystyle{IEEEtran}

\bibliography{IEEEabrv,simple}

\end{document}